\documentstyle[preprint,psfig,aps]{revtex}
\newcommand{\beq}{\begin{equation}}
\newcommand{\eeq}{\end{equation}}
\newcommand{\bdis}{\begin{displaymath}}
\newcommand{\edis}{\end{displaymath}}
\newcommand{\bea}{\begin{eqnarray}}
\newcommand{\eea}{\end{eqnarray}}
\newcommand{\barr}{\begin{array}}
\newcommand{\earr}{\end{array}}

\begin{document}

\title{ Correlations in Economic Time Series}

\author{ Yanhui Liu, Pierre Cizeau, Martin Meyer, Chung-Kang Peng, and
H.~Eugene Stanley} 
\address{Center for Polymer Studies and Department of Physics,  Boston
University, Boston, MA 02215, U.S.A.}

\date{May 28, 1997}

\maketitle

\begin{abstract}
The correlation function of a financial index of the New York stock
exchange, the S\&P 500, is analyzed at 1$\,$min intervals over the
13-year period, Jan 84 -- Dec 96. We quantify the correlations of the
absolute values of the index increment. We find that these correlations
can be described by two different power laws with a crossover time
$t_\times\approx 600\,$min. Detrended fluctuation analysis gives
exponents $\alpha_1=0.66$ and $\alpha_2=0.93$ for $t<t_\times$ and
$t>t_\times$ respectively. Power spectrum analysis gives corresponding
exponents $\beta_1=0.31$ and $\beta_2=0.90$ for $f>f_\times$ and $f<
f_\times$ respectively.

\end{abstract}

\bigskip

A topic of considerable recent interest to both the economics and
physics communities is whether there are correlations in economic time
series and, if so, how to best quantify these correlations
\cite{Mandelbrot63,Fama70,Ding93,Dacorogna93}.  Here we study the S\&P
500 index of the New York stock exchange over a 13-year period
(Fig.~1a).  We calculate the logarithmic increments $g(t)\equiv\ln
Z(t+1) - \ln Z(t)$ over a fixed time lag of 1$\,$min, where $Z(t)$
denotes the index at time $t$ ($t$ counts the number of minutes
during the opening hours of the stock market), and quantify the
correlations as follows:

(i) We find that the correlation function of $g(t)$ decays exponentially with
a characteristic time of the order of 1-10$\,$min, but the absolute value
$|g(t)|$ does not. This result is consistent with previous studies on
several economic series \cite{Fama70,Ding93,Dacorogna93}. 

(ii) We calculate the power spectrum of $|g(t)|$
(Fig.~2a), and find that the data fit not one but rather two separate power
laws: for $f>f_\times$ the power law exponent is $\beta_1=0.31$, while
for $f< f_\times$ the exponent $\beta_2=0.90$ is three times larger;
here $f_\times$ is called the crossover frequency.

(iii) We confirm these results using the DFA (detrended fluctuation analysis)
method  (see Fig.2b), which allows accurate estimates of
exponents {\it independent} of local trends \cite{Peng94}.  From the
behavior of the power spectrum, we expect that the DFA method will also
predict two distinct regions of power law behavior, with exponents
$\alpha_1 = 0.66$ and $\alpha_2 = 0.95$ for $t$ less than or greater
than a characteristic time scale $t_\times \equiv 1/f_\times$, where we
have used the general mathematical result \cite{Beran94} that
$\alpha=(1+\beta)/2$. The data of Fig.~2b yield $\alpha_1=0.66$,
$\alpha_2=0.93$, thereby confirming the consistency of the power
spectrum and DFA methods. Also the crossover time is very close to the
result obtained from the power spectrum, with
$t_\times\approx1/f_\times\approx 600\,$min (about 1.5 trading days). 

We observed the crossover behavior noted above by considering the
entire 13-year period studied, so it is natural to enquire whether it
will still hold for periods smaller than 13 y.  Therefore, we choose a
sliding window (with size 1$\,$y) and calculate both exponents
$\alpha_1$ and $\alpha_2$ within this window as the window is dragged,
down the data set.  We find (Fig.~1b) that the
value of $\alpha_1$ is very ``stable'' (independent of the position of
the window) fluctuating around the mean value 2/3. Surprisingly,
however, the variation of $\alpha_2$ is much greater, showing sudden
jumps when very volatile periods enter or leave the time window.

We studied several standard mathematical models, such as fractional
Brownian motion \cite{Beran94,Mandelbrot68} and fractional ARIMA
processes \cite{Granger96}, commonly used to account for long-range
correlation in a time series and found that none of them can reproduce
the large fluctuation of $\alpha_2$.

\noindent
{\bf Acknowledgments:} We thank S.~Havlin, R.~Mantegna, and S.~Zapperi
for extremely helpful discussions through the course of this work.

\pagebreak

\begin{figure}
\caption{
(a) Raw data analyzed: The S\&P 500 index $Z(t)$for the 13-year period
1 Jan 1984 -- 31 Dec 1996 at intervals of 1 min. Note the large
fluctuations, such as that on 19 Oct 1987 (``black Monday''). (b)
Results of dragging a window of size 1$\,$y down the same data base,
one month at a time, and calculating the best fit exponent $\alpha_1$ 
(dashed line) and $\alpha_2$ (full line) for the time intervals  $t<
t_\times$ and $t>t_\times$ respectively.
}
\end{figure}

\begin{figure} 
\caption{ Plot of (a) the power spectrum $S(f)$ and (b) the detrended
fluctuation analysis $F(t)$ of the absolute values of the 1$\,$min
increments.  The lines show the best power law fits ($r$ values are
better than $0.99$) to the data above and below the crossover frequency
of $f_\times=(1/570)\,$min$^{-1}$ in (a) and of the crossover time
$t_\times=600\,$min in (b). To remove artificial correlations resulting
from the intra-day pattern of the market activity
\protect{\cite{Wood,Harris}}, we analyze normalized data $|g_n(t)|\equiv
|g(t)|/A(t)$, where $A(t)$ is the activity at the same time of the day
averaged over all days of the data set. For the DFA method, we integrate
$|g_n(t)|$ once; then we determine the fluctuations $F(t)$ of the
integrated signal around the best linear fit in a time window of size
$t$.  }
\end{figure}

\thispagestyle{empty}
\vspace*{\fill}
\centerline{\psfig{figure=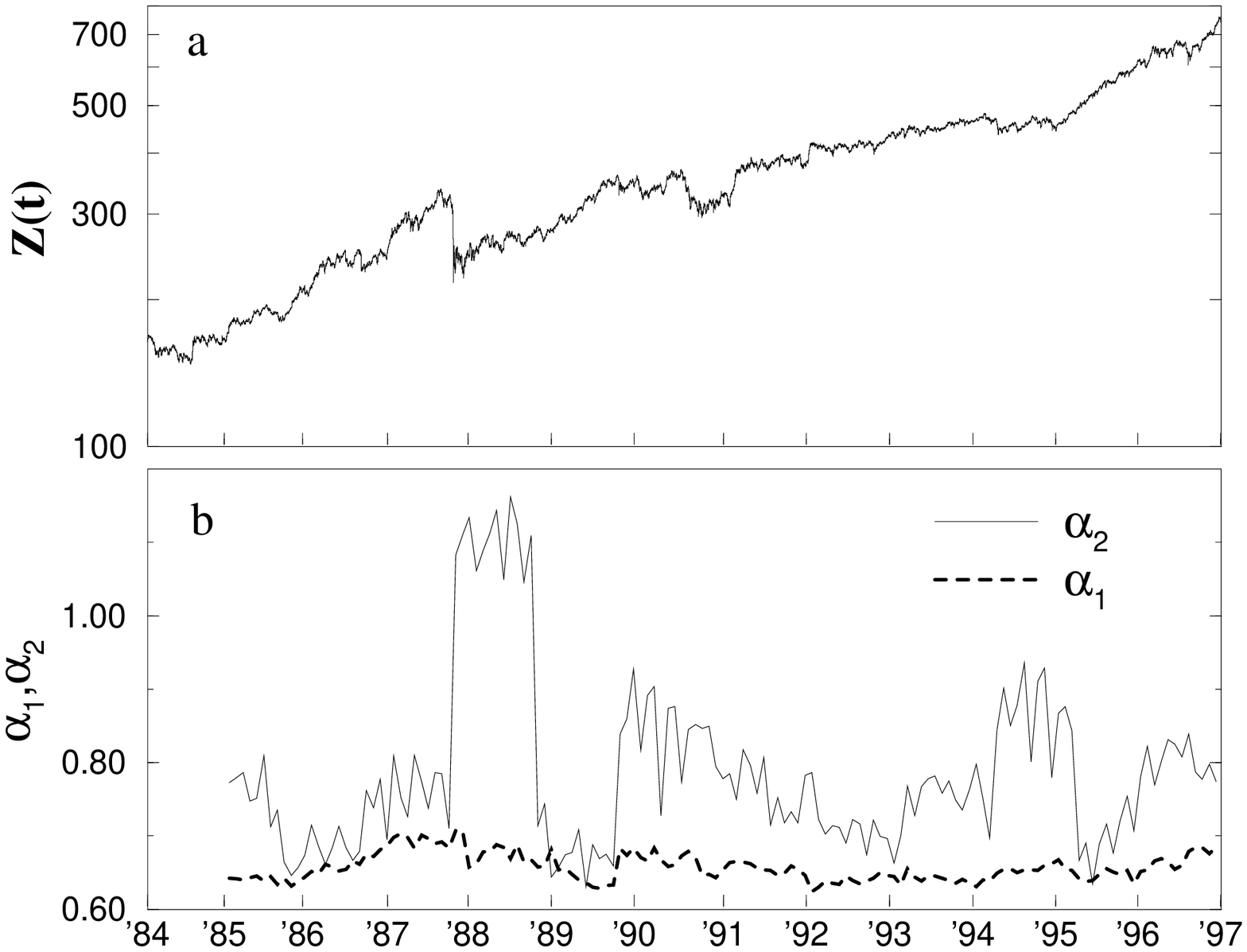,width=6in}}
\vfill
\begin{center}
{\huge  \bf Fig. 1}
\end{center}

\thispagestyle{empty}
\vspace*{\fill}
\centerline{\psfig{figure=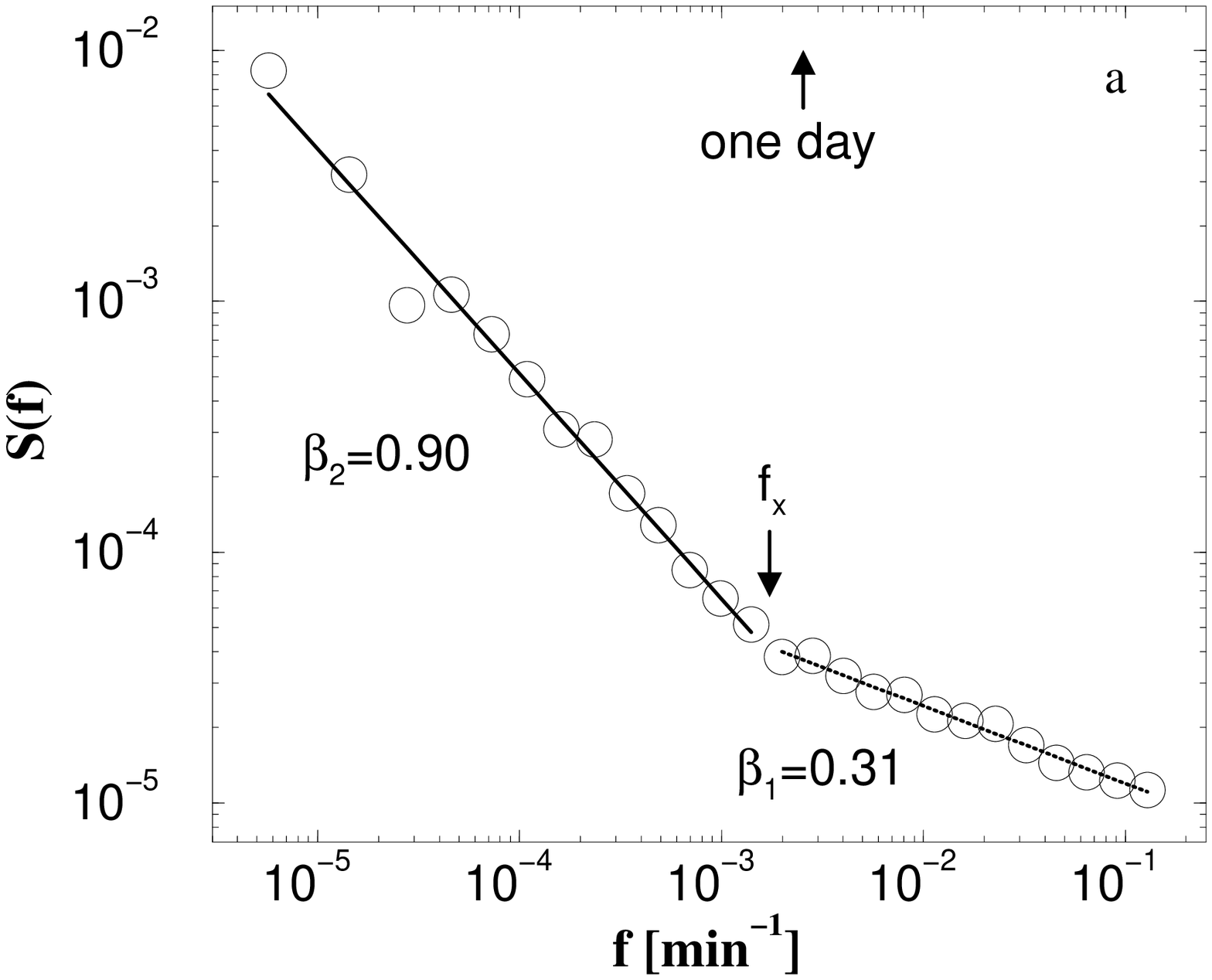,width=4.4in}}
\centerline{\psfig{figure=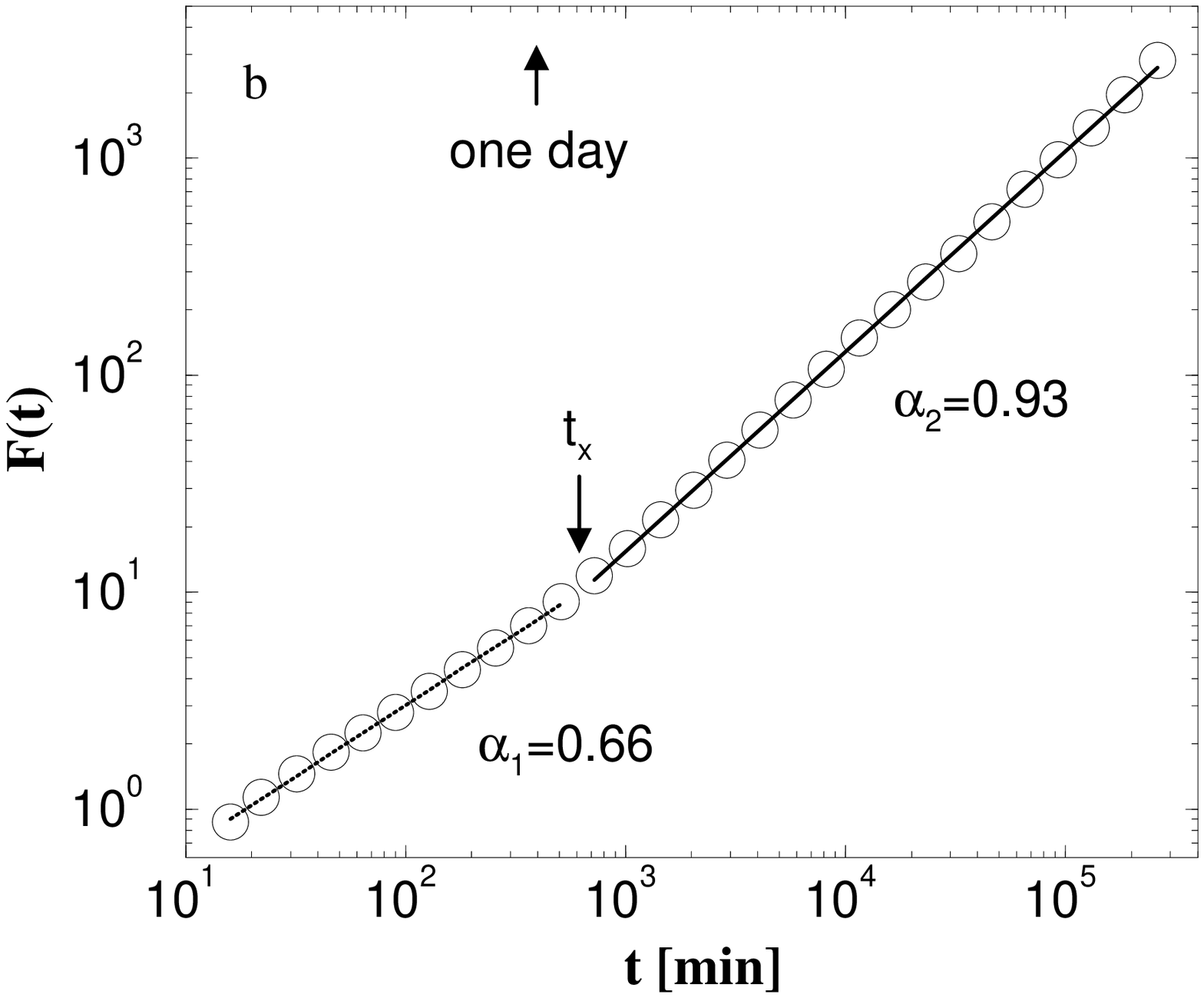,width=4.4in}}
\vfill
\begin{center}
{\huge \bf Fig. 2}
\end{center}

\end{document}